\documentclass[aps,prb,onecolumn,superscriptaddress,floatfix,showpacs,amsmath,amssymb,nofootinbib,longbibliography,nobibnotes,noeprint,11pt]{revtex4-2}

\usepackage{graphicx}
\usepackage{dcolumn,xcolor}
\usepackage{bm}
\usepackage{color}
\usepackage{tabularx}
\usepackage{ragged2e}
\usepackage{siunitx}
\usepackage{verbatim}
\usepackage{parskip,stackengine}
\usepackage{soul,xr}
\usepackage[normalem]{ulem}
\usepackage{hyperref}
\hypersetup{
	pdfpagemode=UseOutlines,
	pdfstartview=Fit,
	pdftoolbar=true,
	colorlinks,
	linkcolor=blue,
	citecolor=blue,
	urlcolor=black,
	filecolor=blue,
	runcolor=blue,
	breaklinks,
	bookmarksnumbered=true,
	bookmarksopen,
	bookmarksopenlevel=1,
	pdftitle={ },
	pdfsubject={ },
	pdfauthor={ },
	pdfkeywords={}}
\newcolumntype{C}[1]{>{\centering\arraybackslash}p{#1}}
\newcolumntype{L}[1]{>{\raggedright\arraybackslash}p{#1}}
\newcolumntype{R}[1]{>{\raggedleft\arraybackslash}p{#1}}
\newcolumntype{J}[1]{>{\justifying\arraybackslash}p{#1}}

\newcommand{\ti}[1]{\textsubscript{#1}}
\newcommand{\te}[1]{\textsuperscript{#1}}

\newcommand{\B}{\ensuremath{B}}

\newcommand{\muB}{\ensuremath{\mu_\text{B}}}
\newcommand{\Bc}{\ensuremath{B_\text{c}}}
\newcommand{\Bs}{\ensuremath{B_\text{s}}}
\newcommand{\Bcone}{\ensuremath{B_\text{c1}}}
\newcommand{\NCTO}{Na\ti{2}Co\ti{2}TeO\ti{6} }
\newcommand{\tripleq}{triple-$\mathbf{q}$ }

\begin{document}

\title{Spin waves in Na$_2$Co$_2$TeO$_6$ studied by high-frequency/high-field ESR:\\ Successes and failures of the triple-$\mathbf{q}$ model}

\author{Luca~Bischof}
\email{luca.bischof@kip.uni-heidelberg.de}
\affiliation{Kirchhoff Institute of Physics, Heidelberg University, 69120 Heidelberg, Germany}

\author{Jan~Arneth}
\affiliation{Kirchhoff Institute of Physics, Heidelberg University, 69120 Heidelberg, Germany}

\author{Raju~Kalaivanan}
\affiliation{Institute of Physics, Academia Sinica, Taipei 11529, Taiwan}

\author{Raman~Sankar}
\affiliation{Institute of Physics, Academia Sinica, Taipei 11529, Taiwan}

\author{Kwang-Yong~Choi}
\affiliation{Department of Physics, Sungkyunkwan University, Suwon 16419, Republic of Korea}

\author{Rüdiger~Klingeler}
\email{klingeler@kip.uni-heidelberg.de}
\affiliation{Kirchhoff Institute of Physics, Heidelberg University, 69120 Heidelberg, Germany}

\date{\today}

\begin{abstract}
The Kitaev candidate material Na$_2$Co$_2$TeO$_6$ is proposed to be proximate to a quantum spin liquid state but a suitable spin model and the nature of its ground states are still under debate. Our high-frequency/high-field electron spin resonance spectroscopy studies of Na$_2$Co$_2$TeO$_6$ single-crystals under in-plane and out-of-plane magnetic fields elucidate the ground state by investigating its low-energy spin wave excitations. Several excitation modes are observed in the low-field phase and in the phases induced by $B\parallel a^*$. In addition, the spectra exhibit a frequency-independent feature at the phase boundary connected to the putative quantum phase transition. For magnetic fields applied along the $c$ axis, the observation of three distinct spin wave modes in the antiferromagnetic (AFM) ground state reveals a previously unresolved splitting of the zero-field excitation gap into $\Delta = 211\,$GHz and $\Delta_2 = 237\,$GHz. The softening of one of these modes evidences a field-induced phase transition at $B_{\rm c1} = 4.7\,$T, which is corroborated by a clear anomaly in the isothermal magnetization. Spin wave calculations based on the extended Heisenberg-Kitaev model exclude a zigzag ground state of the AFM phase. A triple-q spin configuration correctly predicts two spin wave modes, but fails to reproduce the softening mode. Our analysis shows that the triple-q ground state model of Na$_2$Co$_2$TeO$_6$ is incomplete and suggests the relevance of interlayer interactions.

\end{abstract}

\maketitle

\section{Introduction}

\begin{figure*}[ht]
\includegraphics[width=\textwidth,clip]{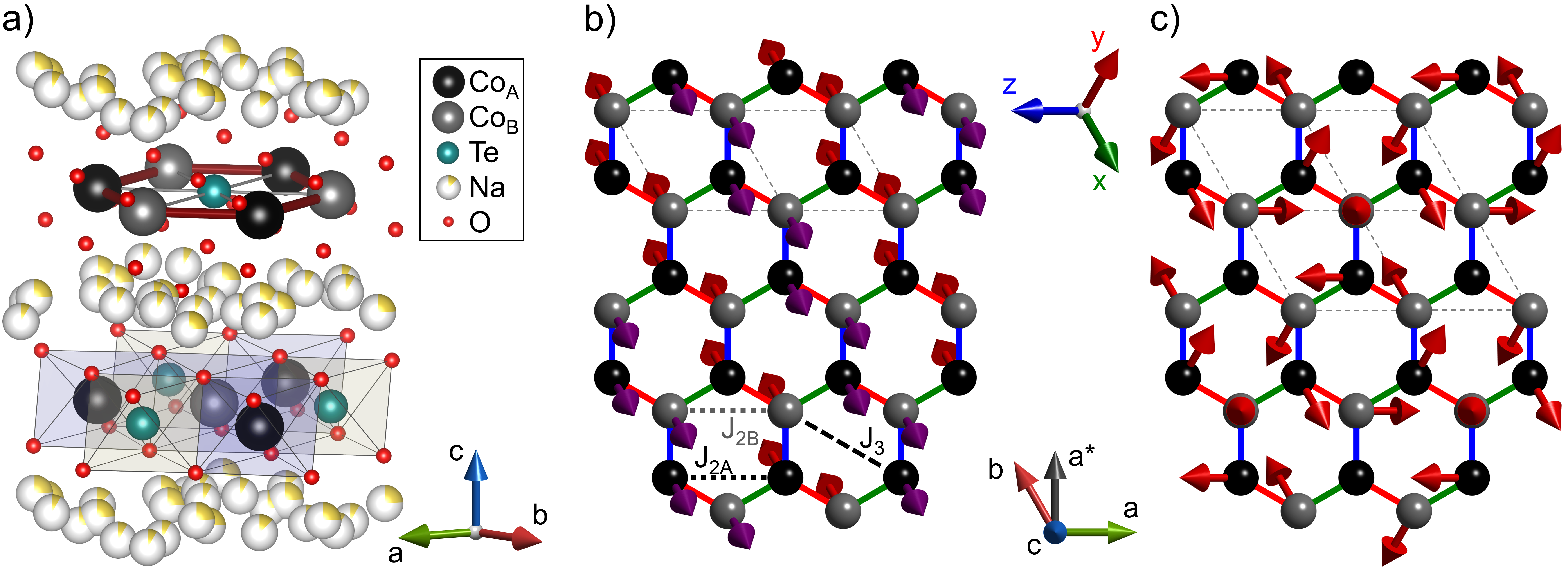}
\caption{Crystal structure and proposed magnetic structures of the ground state of Na\ti{2}Co\ti{2}TeO\ti{6}. (a) Three-dimensional crystal structure consisting of alternating layers of CoO\ti{6} and TeO\ti{6} octahedra separated by partially occupied Na sites.~\cite{Lefrancois2016} Octahedral structure and hexagonal Co-layers are highlighted in the lower and upper layer, respectively. The figure was generated by VESTA.~\cite{Vesta2011} (b,c) In-plane Co-lattice showing the $x$-, $y$-, and $z$-bonds (green, red, and blue, respectively) in the hexagonal structure. Arrows depict the spin orientation of (b) zigzag and (c) \tripleq ground states with their magnetic unit cell (dashed lines), whereas arrows obscured by their ion point antiparallel to the $c$ axis.  Next-neighbor interactions $J_\text{2A}$ and $J_\text{2B}$ on the two Co-sublattices (black, gray), third-neighbor coupling $J_3$, and the spin ($x,y,z$) and lattice ($a,b,c$) coordinate systems are indicated.}
\label{Fig_Intro}
\end{figure*}

\noindent One of the most promising routes for realizing a quantum spin liquid (QSL) was established by the seminal Kitaev model \cite{Kitaev2006}. Its exact solution -- a quantum spin liquid state featuring long-range entanglement yet no long-range order and Majorana fermions as fractionalized elementary excitations -- relies on $S=1/2$ spins on a honeycomb lattice with magnetic frustration induced by bond-dependent Ising interactions 

\begin{equation}\label{eq_H-K}
    \mathcal{H}_\text{K} = \sum_{\langle i,j\rangle^\gamma} K\cdot S^\gamma_i S^\gamma_j 
\end{equation}

\noindent with the three neighboring sites ($\gamma = x,y,z$). Starting the quest for materials realizing the Kitaev model, compounds comprising \textit{d}\te{5} transition metal ions with strong spin-orbit coupling have been studied.~\cite{Jackeli2009,Takagi2019} However, instead of an anticipated QSL ground state, these \textit{d}\te{5} Kitaev candidates exhibit long-range magnetic order, which is attributed to further magnetic interactions superimposed on the Kitaev coupling \eqref{eq_H-K}. Among them, Na\ti{2}IrO\ti{3} \cite{Chun2015,Chaloupka2013} and $\alpha$-RuCl\ti{3} \cite{Do2017,Kasahara2018,Janssen2017} received particular attention because the experimental findings suggest proximity to a QSL state. Later, the search for Kitaev materials was expanded to high-spin $d^7$ ions ($t_{2g}^5 e_g^2$) such as Co\te{2+}~\cite{Liu2018,Sano2018,Liu2020}: Edge-sharing octahedral Co-environments with 90° Co-O-Co bonding geometry lead to Kitaev interactions between the effective $S=1/2$ moments describing the low-energy spin-orbit Kramers doublet, while the spin exchange involving $e_g$ electrons suppresses the Heisenberg interaction.

\NCTO is the prime example of $d^7$ Kitaev candidate compounds due to its remarkable similarity to $\alpha$-RuCl\ti{3} and its predicted proximity to the QSL phase.~\cite{Lin2021,Kruger2023} Although \NCTO fulfills the structural requirements necessary for dominant Kitaev interactions (see Fig.~\ref{Fig_Intro})~\cite{Viciu2007,Lefrancois2016} and can be described by an effective spin $S=1/2$ (hereafter referred to as 'spin')~\cite{Songvilay2020,Yao2022}, it exhibits 3D antiferromagnetic order (AFM) below $T_\text{N} = 26.8(3)\,$K with preceding in-plane ordering at $31\,$K \cite{Arneth2024,Chen2021}. Magnetic fields applied along the armchair direction $a^*$ of the hexagonal lattice suppress $T_\text{N}$ and lead to a field-induced phase above $\Bc=6.1\,$T at low temperatures~\cite{Arneth2024,Kikuchi2022,Zhang2023,Hong2021,Yao2020}, which is of putatively quantum-disordered nature.~\cite{Lin2021} For the phase boundary at $\Bc$, signatures of a quantum critical endpoint were found.~\cite{Arneth2024}

Despite years of intense research on Na\ti{2}Co\ti{2}TeO\ti{6}, even the spin configuration of the AFM ground state at zero magnetic field is still under debate: Based on neutron powder diffraction, the ground state was first reported to be a collinear zigzag structure (Fig.~\ref{Fig_Intro}b) consisting of ferromagnetic zigzag chains with antiparallel spin orientations on neighboring chains.~\cite{Lefrancois2016,Bera2017,Samarakoon2021} Using single-crystal neutron diffraction and inelastic neutron scattering (INS), the ground state was later revised to be a \tripleq configuration (Fig.~\ref{Fig_Intro}c)~\cite{Chen2021}, which can be constructed by superposing three zigzag configurations rotated by 120°. These conflicting results arise from the inability of neutron diffraction to distinguish between multi-domain states (three different orientations of zigzag order) and multi-$\mathbf{q}$ structures such as the \tripleq configuration because both exhibit the same diffraction pattern.~\cite{Chen2021,Paddison2024} Recently, experiments specifically designed to distinguish the proposed ground states~\cite{Yao2023,Jin2025} and theoretical studies~\cite{Kruger2023,Francini2024fM} have supported the notion of a \tripleq ground state in Na\ti{2}Co\ti{2}TeO\ti{6}. However, several other studies using electric polarization, optical magnetospectroscopy, neutron diffraction, and muon-spin rotation interpret their results in favor of a zigzag ground state. \cite{Zhang2023,Xiang2023,Jiao2024}
The search for microscopic models for Na\ti{2}Co\ti{2}TeO\ti{6} in terms of spin Hamiltonians is complicated since it relies substantially on the realized ground state of a material to fit the spin wave spectra probed by INS. This search is further hindered by the high-dimensional parameter space spanned by the Kitaev interaction \eqref{eq_H-K} and several additional exchange couplings.

In this work, we study the low-energy spin excitations in Na\ti{2}Co\ti{2}TeO\ti{6} for in-plane ($\B\parallel a^*$) and out-of-plane ($\B\parallel c$) magnetic fields by high-frequency/high-field electron spin resonance spectroscopy (HF-ESR). The spin wave modes observed across various magnetic phases provide information on the magnetic ground states and the underlying spin Hamiltonian, and reveal a splitting of the zero-field excitation gap at low energies as well as a field-induced phase transition for $\B\parallel c$. By comparing the observed modes with spin wave calculations, we exclude a zigzag ground state in Na\ti{2}Co\ti{2}TeO\ti{6}. Our data further show that the recently proposed \tripleq model only provides an incomplete description of the low-energy spin dynamics such that none of the Hamiltonians reported to date are sufficient to describe the physics governing the magnetism in Na\ti{2}Co\ti{2}TeO\ti{6}.

\section{Experimental Details and Methods}
\label{Sec_Methods}

\begin{figure*}[t]
\includegraphics[width=\textwidth,clip]{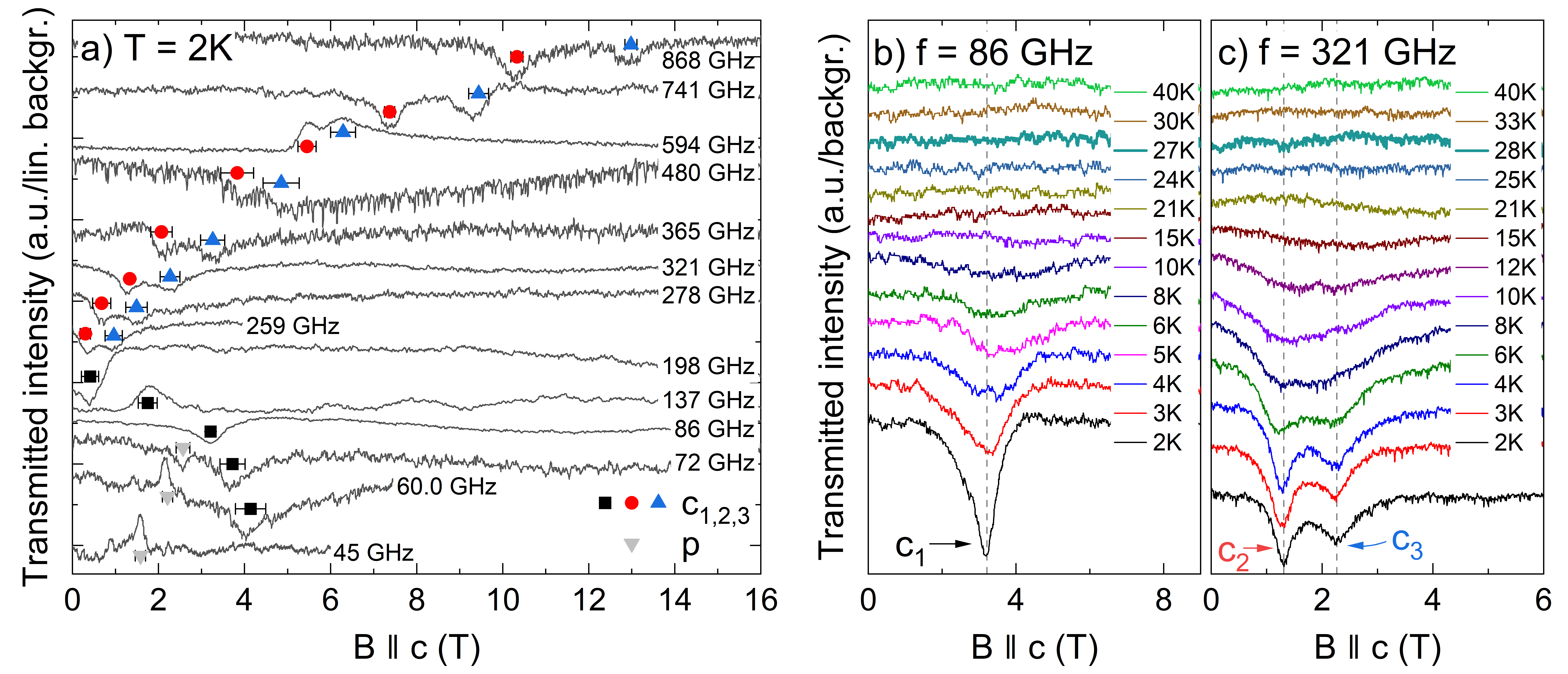}
\caption{HF-ESR spectra of Na\ti{2}Co\ti{2}TeO\ti{6} for out-of-plane magnetic fields $\B\parallel c$. (a) Selected spectra obtained at $T=2\,$K for frequencies ranging from $45\,$GHz to $868\,$GHz. Resonances assigned to the three magnon modes $c_1,c_2$, and $c_3$, and the paramagnetic impurity mode $p$ are marked by symbols. The temperature dependence of the spectra at (b) $f=86\,$GHz and (c) $f=321\,$GHz tracks the evolution of the indicated spin wave modes upon heating. Vertical dashed lines mark the resonance positions at $T = 2\,$K.}
\label{Fig_SpectraC}
\end{figure*}

\noindent
High-frequency/high-field electron spin resonance (HF-ESR) measurements were performed on a high-quality single-crystal of \NCTO grown by the self-flux method.~\cite{Lee2021} A millimeter vector network analyzer (MVNA) from \textit{ABmm} was used as a phase-sensitive microwave source and detector \cite{Comba2015} to obtain transmission spectra at frequencies ranging from $45\,\mathrm{GHz}$ to $870\,\mathrm{GHz}$ in magnetic fields up to $16\,\mathrm{T}$. Temperatures down to $2\,\mathrm{K}$ were achieved in a variable temperature insert (VTI) of an Oxford magnet system.~\cite{Werner2017} The single-crystal sample was mounted in the microwave transmission path with the applied magnetic field parallel to the crystallographic $c$ axis or $a^*$ axis. Spectra were always recorded with both increasing (up-sweep) and decreasing applied magnetic field (down-sweep). All resonances or features identified in a spectrum were -- if not stated otherwise -- observed in both sweep directions. Deviations from the Lorentzian absorption line shape were corrected when determining the resonance fields by using both the amplitude and phase signals. A linear background was subtracted from the spectra at $T=2\,$K, while a phenomenological background correction was employed when measuring the temperature dependency of spectra at fixed microwave frequencies.
Magnetization measurements were performed using the vibrating sample magnetometer option (VSM) of Quantum Design's Physical Properties Measurement System (PPMS-14). 

Spin wave calculations in the presence of an applied magnetic field were performed to compare the experimentally observed magnon frequencies with the predictions of microscopic models of Na\ti{2}Co\ti{2}TeO\ti{6}. Without a magnetic field, the ground state configurations were determined in two steps: First, the classical energy of the respective ground state was minimized under consideration of its symmetry. The obtained configuration was then further refined by iterative rotations of each spin toward the local magnetic field.
For zigzag ground states with identical spin orientation along zigzag chains and opposite orientation on neighboring zigzag chains (Fig.~\ref{Fig_Intro}b), two independent spherical angles determining the configuration were optimized during the first step. 
The triple-$\mathbf{q}$ order, constructed as a rescaled superposition of three identical but 120° rotated zigzag states, features vortex hexagons separated by spins (anti)parallel to the  $c$ axis (Fig.~\ref{Fig_Intro}c). The six spins on vortex hexagons are related to their neighbors by a 60° rotation and an inversion of the out-of-plane spin component. The triple-$\mathbf{q}$ order is thus defined by two spherical angles, which were varied to minimize energy.
The in-field ground states were obtained by incrementally altering the applied magnetic field from the zero-field ground configuration or known in-field states and refining the spin configuration after each field increment. The obtained spin states were cross-checked at selected magnetic fields with Monte Carlo sampling in the full configuration space of the $2\times1$ ($2\times2$) magnetic unit cell with 4 (8) spins for zigzag (triple-$\mathbf{q}$) states and subsequent refinement. Based on the resulting ground state spin configurations, the spin wave frequencies were computed in the framework of linear spin wave theory.~\cite{Fishman2018} Calculations were performed using the \textit{SpinW} library.~\cite{Toth2015}

\section{Results}
 
\subsection{Spin waves in \NCTO for $\mathbf{B\parallel c}$}

\begin{figure}[]
\includegraphics[width=\columnwidth,clip]{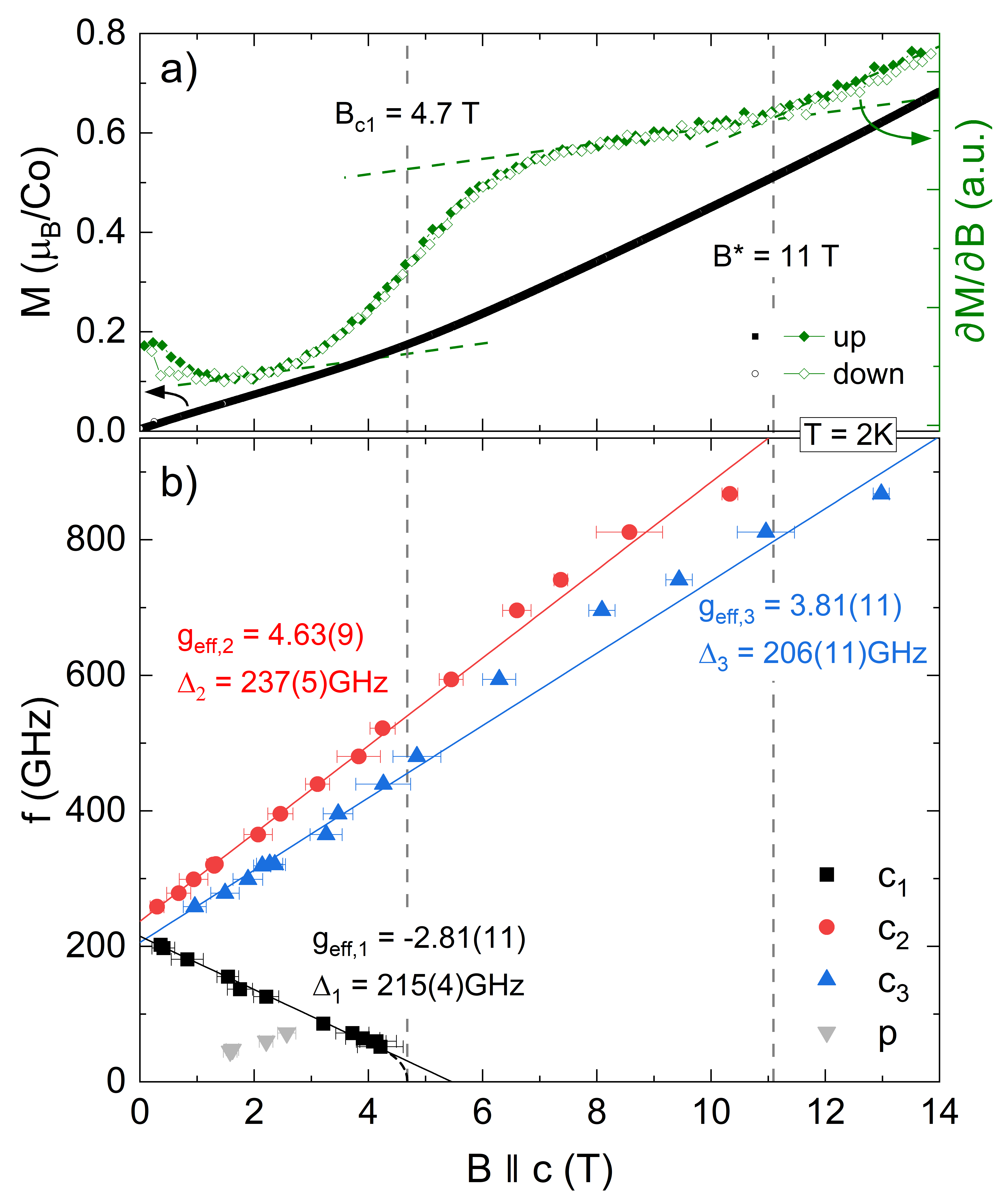}
\caption{Spin waves and magnetization for out-of-plane magnetic fields $\B\parallel c$. (b) Resonance frequency-field diagram at $T=2\,$K. The resonances assigned to the three spin wave modes $c_i$ are fitted  by $hf = \Delta_i + g_{\text{eff},i}\times \muB B$ (solid lines) with effective $g$-factors and excitation gaps $\Delta_i$ as stated. The mode $c_1$ is non-linearly extrapolated (dashed line) such that its softening field  (vertical line) coincides with a kink in the magnetization $M$ at $\Bcone =4.7\,$T [(a), left axis], i.e., a step-like increase in the magnetic susceptibility $\partial M/\partial B$ (right axis) shown for increasing (up) and decreasing (down) magnetic fields.}
\label{Fig_C_ffdiag}
\end{figure}

\begin{figure*}
    \includegraphics[width=\textwidth,clip]{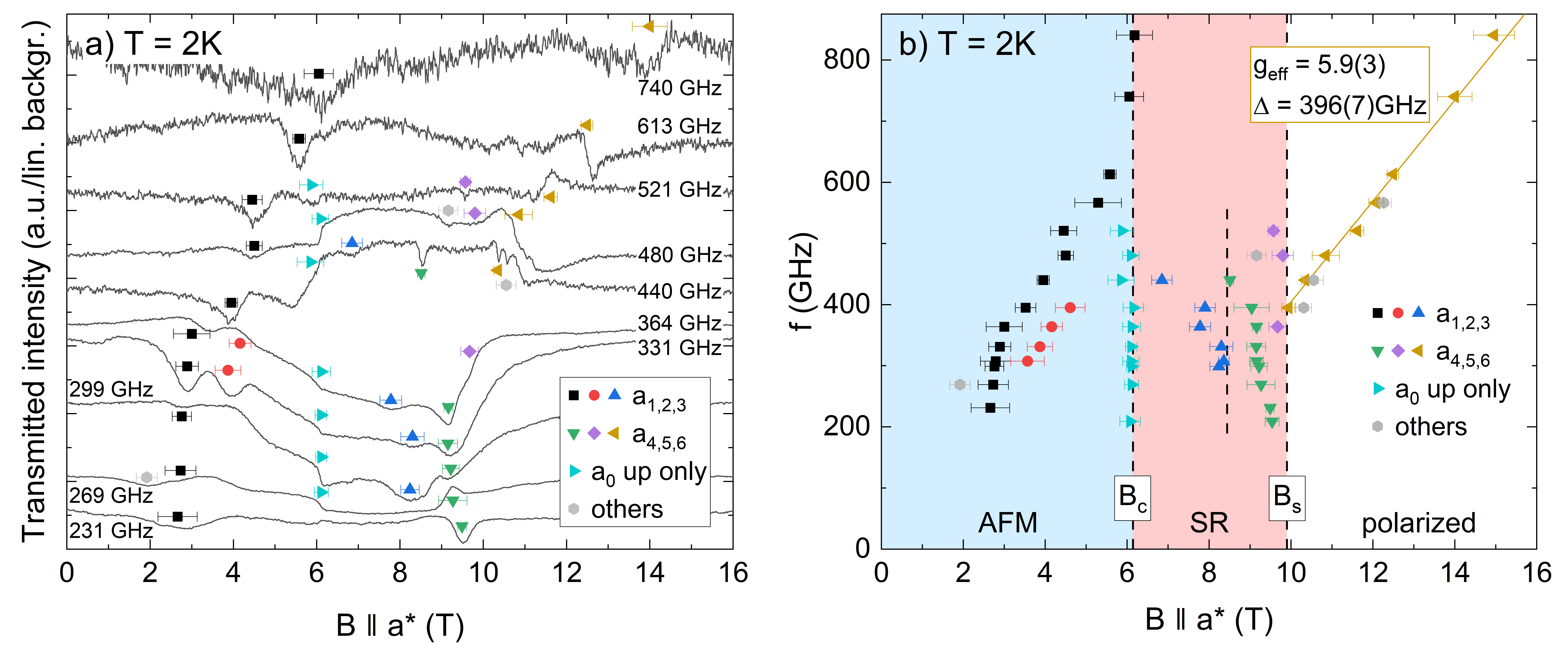}
    \caption{HF-ESR of Na\ti{2}Co\ti{2}TeO\ti{6} for in-plane magnetic fields $\B\parallel a^*$ at $T=2\,$K. (a) Spectra at selected frequencies from $231\,$GHz to $740\,$GHz. 
    The identified features are grouped as marked by different symbols and shown in the frequency-field diagram (b). The feature $a_0$ is only observed for increasing applied magnetic field. The transition to the field-induced intermediate phase, at    
    $\Bc$, and the saturation field $\Bs$ are marked by vertical lines which separate the antiferromagnetically ordered phase (AFM), the spin-reoriented phase (SR), and the polarized region as color-coded. The vertical line separating $a_3$ and $a_4$ in the SR phase is a guide to the eye. The spin wave mode $a_6$ is fitted by $hf = \Delta + g_\text{eff}\cdot \muB (B-\Bs)$ (see the text).}
    \label{Fig_Spectra-ff-diag_As}
\end{figure*}

\noindent
HF-ESR spectra of \NCTO for out-of-plane magnetic fields $\B\parallel c$ are shown in Fig.~\ref{Fig_SpectraC}a. The measurements were performed at $T=2\,$K, which is well below the long-range magnetic ordering temperature $T_\text{N}=27\,$K. Distinct resonances are observed and assigned to excitation modes $c_1,c_2$, and $c_3$ persisting in the spectra over a wide range of microwave frequencies. For $f\geq 259\,$GHz, the resonance positions of the modes $c_2$ and $c_3$ shift toward higher fields with increasing frequency. In contrast, upon reducing $f$ below $259\,$GHz, the resonance mode $c_1$ is found at increasing magnetic fields. In addition, a narrow feature $p$ is observed at a few frequencies below $72\,$GHz. Deviations from the Lorentzian absorption line shape at several frequencies are caused by wave-phase mixing~\cite{Jonak2022} and are corrected when determining the resonance fields.

All observed resonance features found for $\B\parallel c$ are summarized in the frequency-field diagram in Fig.~\ref{Fig_C_ffdiag}b. We clearly identify three main resonance modes $c_1,c_2$, and $c_3$, which correspond to the spin wave excitations (magnons) of the antiferromagnetically ordered phase of \NCTO at low temperatures. Starting from similar zero-field excitation energies, the resonance frequency of $c_1$ decreases while the frequencies of $c_2$ and $c_3$ increase with magnetic field. The frequency-field dependencies are fitted by the gapped linear relation $hf = \Delta_i + g_{\text{eff},i}\times \muB B$, which is used to determine the zero-field excitation gaps $\Delta_i$ and the effective \textit{g}-factors $g_{\text{eff},i}$ of the three branches (see Fig.~\ref{Fig_C_ffdiag}b and Tab.~\ref{Tab_g-eff}). Extrapolating this linear behavior to $B=0\,$T shows that \NCTO exhibits a split excitation gap structure at low energies: The zero-field gaps $\Delta_1=215(4)\,$GHz and $\Delta_3=206(11)\,$GHz of $c_1$ and $c_3$, respectively, agree within their experimental errors, while $\Delta_2=237(5)\,\text{GHz}~\hat{=}~0.98(2)\,$meV is distinguishably larger. For the common zero-field excitation gap $\Delta $ of $c_1$ and $c_3$, we obtain 

\begin{equation*}
    \Delta = \frac{\Delta_1+\Delta_3}{2} = 211(6)\,\text{GHz}~(\hat{=}~  0.87(2)\,\text{meV}).
\end{equation*}

\noindent In contrast to the two distinct zero-field excitation gaps $(\Delta,\Delta_2)$ revealed in this work, single-crystal INS~\cite{Chen2021,Yao2022} and a previous single-crystal HF-ESR~\cite{Bera2023} study resolved only one broad low-energy zero-field excitation at energies similar to $\Delta$ and $\Delta_2$. A much smaller gap was obtained by HF-ESR on polycrystalline samples~\cite{Lin2021}. Our data cannot exclude flattening of the frequency-field dependencies of $c_1,c_2$, and $c_3$ close to zero magnetic field. Such a flattening, albeit not expected by theory (see Sec.~\ref{Sec_SpinWave_3q}), could separate the gaps of $c_1$ and $c_3$, which agree for linear extrapolation, but preserves the split gap structure by even increasing the difference between $\Delta_1$ and $\Delta_2$.~\cite{Kurbakov2017} Hence, our experimental data unambiguously prove the existence of (at least) two spin wave excitation gaps up to $900\,$GHz.

With increasing magnetic field, the resonance frequency of the spin wave mode $c_1$ decreases until no resonances are observed below $52\,$GHz. A linear extrapolation of this resonance branch predicts the softening of $c_1$ near $5\,$T. Such a softening is usually connected with an instability of the magnetic order and indicates a magnetic field-induced phase transition. At the same field, the spin wave modes $c_2$ and $c_3$ start to deviate from their linear frequency-field dependence. Notably, while $c_1$ softens near $5\,$T and is no longer observed at larger magnetic fields, the frequencies of $c_2$ and $c_3$ increase continuously in this field regime. For $c_3$, the data do not exclude the presence of a small jump in the resonance frequency at $5\,$T. 
Concomitant with the softening of $c_1$ and the changes in $c_2$ and $c_3$, the material features an anomaly of the isothermal magnetization $M$ at $T=2\,$K for out-of-plane magnetic fields (Fig.~\ref{Fig_C_ffdiag}a): $M(B\parallel c)$  exhibits a kink at $\Bcone = 4.7\,$T, which corresponds to a step-like increase in the differential magnetic susceptibility $\partial M/\partial B_{||c}$.~\footnote{$\Bcone $ is determined by the steepest increase in $\partial M/\partial B_{||c}$, which reflects half of the corresponding step-size.} Notably, at \Bcone\ there is also an anomaly in the lattice response as shown by recent magnetostriction data.~\cite{Arneth2025}
The critical field $\Bcone  = 4.7\,$T suggests the non-linear softening of $c_1$ as illustrated by the dashed line in Fig.~\ref{Fig_C_ffdiag}b, which is a typical feature of spin wave modes.~\cite{Turov1965}  We also note that a previous HF-ESR study reported a slightly lower, linearly extrapolated softening field of $4.0\,$T.~\cite{Bera2023}
Both coinciding observations from HF-ESR and static magnetometry, i.e., the softening of the spin wave mode $c_1$ and the increase in susceptibility, strongly suggest a field-driven change of the magnetic structure~\cite{Werner2019}.
 For higher magnetic fields, $\partial M/\partial B_{||c}$ shows a linear field-dependence with only very small slope. We further note that additional left-bending in $\partial M/\partial B_{||c}$ appears for $B>B^* \simeq 11\,$T (Fig.~\ref{Fig_C_ffdiag}a). 

The temperature dependence of the resonance features $c_1$ and $c_2,c_3$ at exemplary frequencies is displayed in Figs.~\ref{Fig_SpectraC}b and \ref{Fig_SpectraC}c, respectively. For all features, a strong decrease in intensity along with line broadening is observed when the temperature increases from $2\,$K. The resonances cannot be distinguished from experimental noise at $T\geq 10\,$K ($T\geq 15\,$K) for $c_1$ ($c_2,c_3$). These temperatures are substantially below the ordering temperature $T_\text{N}=27\,$K, indicating spin wave damping that occurs prior to the complete melting of long-range antiferromagnetic order. Strong spin-phonon scattering~\cite{Hong2024}, unconventional spin dynamics and spin fluctuations at intermediate temperatures below $T_\text{N}$~\cite{Miao2024,Kikuchi2022}, and two-magnon processes~\cite{Kim2021} have been reported for \NCTO and might cause the observed damping.
Finally, we note that the additional resonances $p$ observed in some spectra at low frequencies (Fig.~\ref{Fig_SpectraC}a) are ascribed to parasitic paramagnetic spins due to their gapless linear frequency-field dependence (see Fig.~\ref{Fig_C_ffdiag}b) with a \textit{g}-factor of $2.01(3)$. The different temperature dependence of the resonances $p$ compared to the spin wave modes $c_1,c_2$, and $c_3$ (see Fig.~S1 in the SM~\cite{SM}) corroborates this assignment.

\subsection{HF-ESR for in-plane magnetic fields $\mathbf{\B\parallel a^*}$ }

For magnetic fields applied along the armchair direction $a^*$ of the hexagonal lattice, the HF-ESR spectra of \NCTO at $T=2\,$K exhibit a feature-rich structure. In the frequency range $209\,\text{GHz} \leq f\leq 740\,\text{GHz}$, up to seven features, labeled as $a_i$ ($i=0..6)$, are observed as indicated in Fig.~\ref{Fig_Spectra-ff-diag_As}a. The step-like, frequency-independent features $a_0$ around $6\,$T are only detected for increasing magnetic fields (see Fig.~S3 in the SM~\cite{SM}). No discernible resonance features are observed in the spectra for microwave frequencies below $209\,$GHz.

All features identified for $\B\parallel a^*$ are collected in the frequency-field diagram (Fig.~\ref{Fig_Spectra-ff-diag_As}b). For this field direction, two phase transitions~\cite{Arneth2024,Lin2021} divide the experimentally accessed magnetic field range into three regions: the antiferromagnetic phase (AFM) at low fields up to $\Bc=6.1\,$T, the spin-reoriented phase (SR) at intermediate fields, and the polarized phase above the saturation field $\Bs=9.9\,$T. In the AFM phase, the main resonance $a_1$ exhibits an almost vertical slope at low frequencies, flattens at intermediate frequencies, and eventually diverges toward the phase transition at $\Bc$. A second mode $a_2$ with linear frequency-field relation is observed in the limited frequency range $300\,\text{GHz}< f< 400\,\text{GHz}$. 

The frequency-independent features $a_0$ are straightforwardly associated with the phase transition to the SR phase. This is further confirmed by their temperature dependence, which matches the reported phase boundary $\Bc(T)$ (Figs.~S5 and S6 in the SM~\cite{SM}). The appearance of $a_0$ only in the spectra obtained in increasing magnetic field agrees with the reported magnetic hysteresis~\cite{Arneth2024,Zhang2023} at $\Bc$. In contrast, the resonances $a_1, a_2$ in the AFM phase are observed identically during both increasing and decreasing field sweeps (Fig.~S3 in the SM~\cite{SM}). This suggests that the hysteresis region does not persist down to zero field. 

In the SR phase, two resonance modes $a_3,a_4$ are observed, both exhibiting a decrease in frequency for increasing magnetic field. The behavior of $a_4$ indicates softening at $\Bs$ as expected for the energetically lowest spin wave mode close to the polarized state. However, $a_4$ is not observed at low frequencies in the very vicinity of $\Bs$. Taking a broader look at the spectra shows a region of different microwave transmittance between $6\,$T and $10\,$T with reduced (increased) intensity from $250\,$GHz to $400\,$GHz ($400\,$GHz to $500\,$GHz). This region coincides with the SR phase. We attribute the altered transmittance to different dielectric properties of \NCTO in the SR phase compared to the low- and high-field states \cite{Zhang2023}.
In Refs.~\cite{Hong2021,Takeda2022,Zhang2023}, an additional transition within the SR phase near $8\,$T is discussed. This transition is consistent with the behavior of the modes $a_3,a_4$ observed in this work, which can be understood as a discontinuous change of the resonance frequency at the transition field. 

In the polarized phase, a single spin wave mode $a_6$ is observed, which exhibits a linear frequency-field dependence starting from a finite excitation energy at $\Bs$. Fitting this mode by $hf = \Delta + g_\text{eff}\cdot \muB (B-\Bs)$ yields $\Delta = 396(7)\,$GHz and the effective \textit{g}-factor $g_\text{eff}=5.9(3)$, which is significantly larger than the in-plane \textit{g}-factor $\textit{g}_{ab}=4.13$ (HF-ESR at $T=50\,$K \cite{Lin2021}) expected for the high-field limit. This discrepancy shows that the magnetic interactions in \NCTO significantly contribute to the spin dynamics even in the polarized state in magnetic fields up to $16\,$T. In a THz-spectroscopy study of Na\ti{2}Co\ti{2}TeO\ti{6}, a corresponding mode with a similar $\Delta$ but a smaller effective $\textit{g}$-factor was found \cite{Pilch2023}. Tab.~\ref{Tab_g-eff} summarizes the excitation gaps/energies and effective $g$-factors of all spin wave modes with linear frequency-field dependence observed in this work.

\begin{table}[b]
    \centering
    \def\arraystretch{1.2}
    \caption{Effective \textit{g}-factors and excitation gaps/energies $\Delta$ of the spin wave modes with linear frequency-field dependence. $\Delta$ denotes the spin wave frequency at zero magnetic field (i.e., the zero-field gap) for $\B\parallel c$ and at the phase transition field $\Bs$ for $\B\parallel a^*$, see Fig.~\ref{Fig_C_ffdiag} and Fig.~\ref{Fig_Spectra-ff-diag_As}b.}
    \begin{tabularx}{\columnwidth}{C{0.14\columnwidth}|C{0.14\columnwidth}|C{0.33\columnwidth}|C{0.33\columnwidth}}
                                   & mode     & $g_\text{eff}$                  & $\Delta$ (GHz)  \\  \hline
                                   & $  c_1 $ & $-2.81(11) $                    & $215(4)\phantom{1}$        \\
     $ \B\parallel c\phantom{^*} $ & $c_2$    & $\phantom{-}4.63(9)\phantom{1}$ & $237(5)\phantom{1}$        \\
                                   & $c_3$    & $\phantom{-}3.81(11)$           & $206(11)$       \\  \hline
     $ \B\parallel a^* $           & $a_6$    & $\phantom{-}5.9(3)\phantom{11}$ & $396(7)$    
    \end{tabularx}
    \label{Tab_g-eff}
\end{table}

\section{Spin wave calculations}

\noindent
The spin wave spectrum of an ordered $S=1/2$ spin system is determined by both the magnetic interactions between spins and the realized ground state. Our HF-ESR results thus provide information about the microscopic magnetic properties of \NCTO and allow the testing of theoretical models of its magnetic interactions and the ground state. Spin models for \NCTO~\cite{Songvilay2020,Lin2021,Samarakoon2021,Kim2021,Sanders2022,Kruger2023} are typically based on the extended Heisenberg-Kitaev Hamiltonian
\begin{equation}\label{eq-HKHamiltonian}
    \begin{split}
        \mathcal{H}_\text{HK}=& \sum_{\langle i,j\rangle_{1}^\gamma} \Big[ 
            J_1\big(\mathbf{S}_i\cdot\mathbf{S}_j\big)  
            +K S^\gamma_i S^\gamma_j 
            +\Gamma \big(S^\alpha_i S^\beta_j+S^\beta_i S^\alpha_j\big) \\[-0.1cm]	
        &\hspace{1.4cm}+\Gamma'\big(S^\alpha_i S^\gamma_j+S^\gamma_i S^\alpha_j+ S^\beta_i S^\gamma_j+ S^\gamma_i      S^\beta_j\big)\Big]\\[0.2cm]
        +&\sum_{\langle i,j\rangle_{2}^\Lambda}  J_{2\Lambda}\big(\mathbf{S}_i\cdot\mathbf{S}_j\big)
            +\sum_{\langle i,j\rangle_{3}}  J_{3}\big(\mathbf{S}_i\cdot\mathbf{S}_j\big)
    \end{split}
\end{equation}
for the effective spin $S=1/2$ of the ground state Kramers doublet.
On the level of nearest neighbors, $\mathcal{H}_\text{HK}$ contains all symmetry-allowed bilinear interactions \cite{Chaloupka2015,Francini2024fM}: The isotropic Heisenberg coupling $J_1$, the bond-dependent Kitaev interaction $K$, and symmetric off-diagonal exchanges $\Gamma$ and $\Gamma'$. Nearest neighbor $\gamma$-bonds (Fig.~\ref{Fig_Intro}) are denoted as $\langle i,j\rangle_{1}^\gamma$ with $(\alpha,\beta,\gamma)$ being cyclic permutation of the spin coordinates  $(x,y,z)$. For next-nearest neighbors, isotropic interaction $J_{2\Lambda}$ is included, which can differ between the two crystallographic inequivalent sublattices $\Lambda = \text{A,B}$ \cite{Yao2020,Francini2024fM} marked by gray and black ions in Fig.~\ref{Fig_Intro}. Third-nearest neighbors are coupled by isotropic exchange $J_3$. Further expanding \eqref{eq-HKHamiltonian}, interlayer coupling was considered in Ref.~\cite{Samarakoon2021}. However, its magnitude was reported to be small and it is thus neglected in this work. For applied magnetic fields, a Zeeman term with anisotropic \textit{g}-tensor containing the experimentally determined values $g_c = 2.3$ and $g_{ab} = 4.13$ \cite{Lin2021} is included.
For details of our calculation and the models used, see the appendix~\ref{Sec_Appendix}.

For Na\ti{2}Co\ti{2}TeO\ti{6}, multiple parameter sets for $\mathcal{H}_\text{HK}$ have been proposed in the literature \cite{Songvilay2020,Lin2021,Kim2021,Samarakoon2021,Sanders2022,Kruger2023}. These models are primarily based on INS experiments and were determined by fitting zero-field INS data with calculated spin wave spectra. Conversely, the HF-ESR results obtained in this work reveal information about the spin excitations at vanishing momentum ($\Gamma$-point) in an applied magnetic field. This information is complementary to the INS data and therefore uniquely suitable for testing the proposed models.
For out-of-plane magnetic fields $\B\parallel c$, we observe the three distinct spin wave modes shown in Fig.~\ref{Fig_C_ffdiag}b. They exhibit easy-to-follow frequency-field dependencies and are further detected down to low magnetic fields, making them especially relevant signatures of the spin system in close proximity to the zero-field ground state. In the following, we investigate the low-energy spin wave excitations of extended Heisenberg-Kitaev models for \NCTO with zigzag and triple-$\mathbf{q}$ ground states to compare these to the three modes for $\B\parallel c$. 

\subsection{Zigzag ground state models} \label{Sec_SpinW_Zigzag}

\begin{figure}[]
\includegraphics[width=\columnwidth,clip]{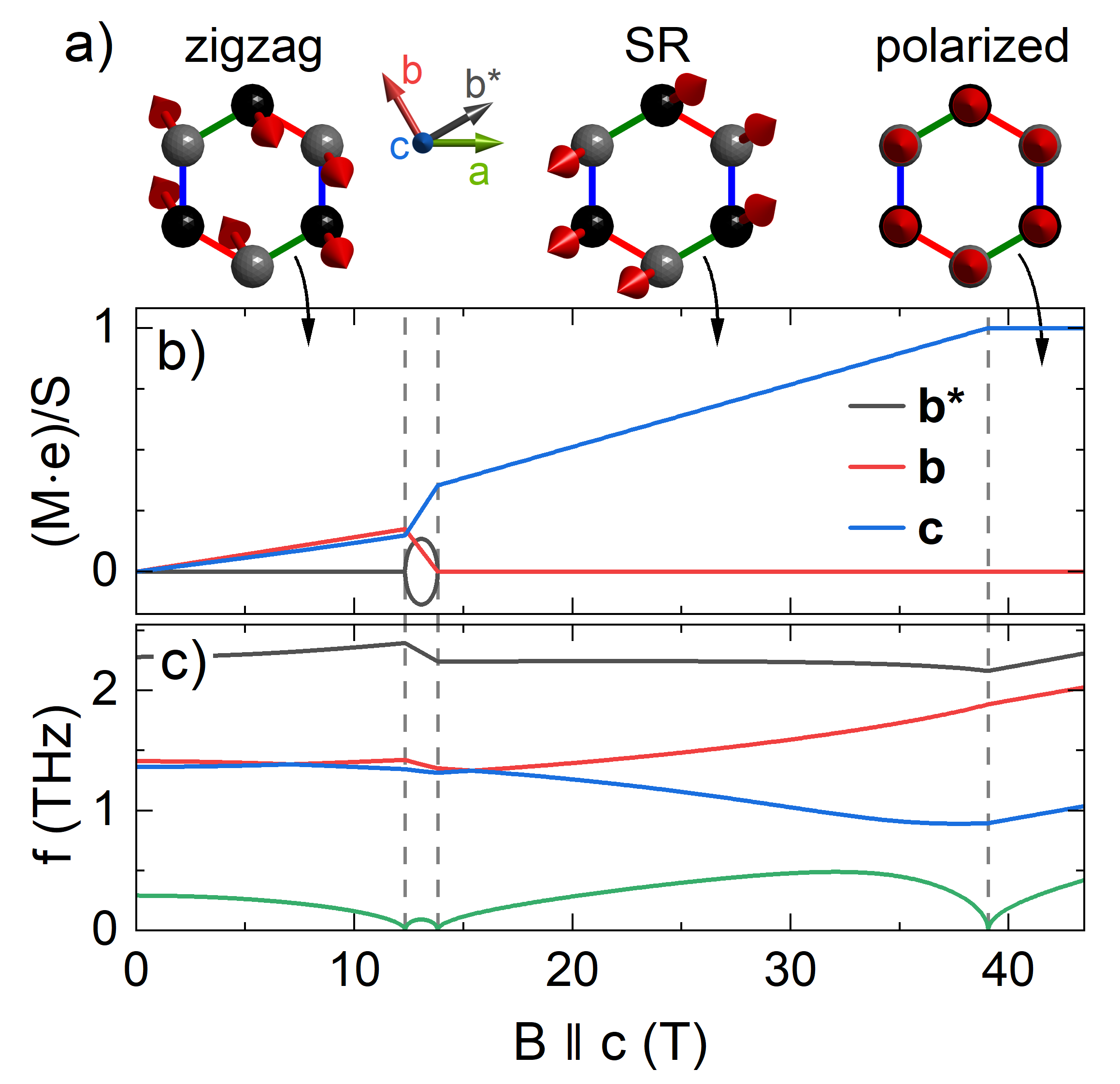}
\caption{Characteristics of the zigzag state in the extended Heisenberg-Kitaev model tx+ for out-of-plane magnetic fields $\B\parallel c$. (a) The spin structure of the ground states at different magnetic fields are illustrated on one hexagon of the honeycomb structure. (b) Normalized total magnetization projected on the indicated crystallographic axes and (c) spin wave frequencies in dependence of the applied magnetic field. Vertical lines mark the transition fields between the zigzag phase, the intermediate phase, the spin-reoriented phase (SR), and the polarized state.}
\label{Fig_SpinWZigzag}
\end{figure}

\noindent
Most of the parameter sets reported for the spin Hamiltonian $\mathcal{H}_\text{HK}$ of \NCTO rely on a zigzag ground state. Exemplarily, the detailed field-dependent behavior of the zigzag ground state model tx+ reported by Sanders~\textit{et al.}~\cite{Sanders2022} is shown for out-of-plane magnetic fields $\B\parallel c$ in Fig.~\ref{Fig_SpinWZigzag}. The interaction parameters of the tx+ model are $(J_1,K,\Gamma,\Gamma') = (-3.5, 3.2, -3, 2)\,$meV, $J_{2A} = J_{2B} = 0$ and $J_3 = 1.4\,$meV (see Tab.~\ref{Appendix_Tab_H-HK-parameter} in the appendix). The classical zigzag ground states at zero magnetic field exhibit ferromagnetic chains along one of the three zigzag directions of the honeycomb lattice ($\mathbf{a},\mathbf{b}$, or $\mathbf{a}+\mathbf{b}$), while spins on neighboring chains are antiparallel. The spins are canted by an angle of 51° out of the honeycomb planes. In the following, we discuss one of these three degenerate ground states, which are connected by a three-fold rotation: the $x$-zigzag ground state (Fig.~\ref{Fig_SpinWZigzag}a) with ferromagnetic chains along the $b$ axis and antiparallel spins on $x$-bonds (green). Its magnetic field dependence calculated in the model by Sanders~\textit{et al.}~\cite{Sanders2022} is illustrated in Fig.~\ref{Fig_SpinWZigzag}: With increasing magnetic field $\B\parallel c$, the spins cant toward $\B$, which leads to a total magnetization with components along the $c$ axis and $b$ axis (Fig.~\ref{Fig_SpinWZigzag}b). At $B=12.3\,$T, the spins begin to rotate and reach a spin-reoriented configuration (SR) at $13.8\,$T, in which all spins cant uniformly toward the field but exhibit opposite components along the $b^*$ axis. Further canting with increasing magnetic field leads to complete polarization at $39.1\,$T. In the intermediate field phase between $12.3\,$T and $13.8\,$T, the clockwise rotation of spins leads to a negative magnetization projection on the $b^*$ axis and to the SR configuration shown in Fig.~\ref{Fig_SpinWZigzag}a with spins pointing outward from the depicted hexagon. Alternatively, an anti-clockwise rotating spin state exhibits the same energy but a positive magnetization component along the $b^*$ axis and leads to an SR configuration with spins pointing into the hexagon, i.e., exchanged spin configuration on the neighboring zigzag chains. Notably, the calculated magnetization along the $c$ axis (blue in Fig.~\ref{Fig_SpinWZigzag}b) qualitatively agrees with out-of-plane magnetization measurements up to $60\,$T \cite{Zhang2024}.

We have calculated the spin wave frequencies in dependence of the magnetic field $\B\parallel c$ based on the ground states of the tx+ model (Fig.~\ref{Fig_SpinWZigzag}c). The number of four modes is determined by the number of magnetic sublattices (two inequivalent sites on two zigzag chains with different spin configurations). We find that the lowest-energy mode softens at the magnetic fields corresponding to the three phase transitions: from the zigzag, through the intermediate, to the SR, and finally the polarized phase. 

To capture the spin waves of zigzag ground states in the Heisenberg-Kitaev model in a more comprehensive picture, we take all proposed models based on a zigzag spin configuration \cite{Songvilay2020,Lin2021,Kim2021,Samarakoon2021,Sanders2022} into consideration. The models' interaction parameters, which in part strongly deviate from each other, are summarized in Tab.~\ref{Appendix_Tab_H-HK-parameter}. We calculated the spin wave spectra for all these models, as shown in Fig.~\ref{Appendix_SpinWaveCalc_zigzag} and discussed in the appendix~\ref{Sec_Appendix}.
In the zigzag phase at low magnetic fields -- on which we focus in the following -- the spin wave spectra of all these models share three characteristic properties with the tx+ model: Firstly, either one or two spin wave modes exist at low energies while further modes are restricted to frequencies $\geq 1\,$THz. Secondly, all spin wave modes exhibit a flat frequency-field dependence at low magnetic fields. At zero field, the corresponding effective \textit{g}-factors $|g_\text{eff}| = |\tfrac{\text{d}f}{\text{d}B}\times h/\muB|\leq 0.012$ are close to zero. Lastly, the softening of the low-lying mode when leaving the zigzag phase toward higher fields occurs at magnetic fields larger than $10\,$T. All of these characteristics are in stark contrast to our HF-ESR data. As shown in Fig.~\ref{Fig_C_ffdiag}b, our results unambiguously reveal three low-energy spin wave modes, strong field-dependence of all observed excitation energies, and a softening of one mode around $5\,$T. Our data, therefore, definitively exclude the reported zigzag ground state models for Na\ti{2}Co\ti{2}TeO\ti{6}.

\subsection{Triple-$\mathbf{q}$ ground state} \label{Sec_SpinWave_3q}

\noindent
The later-proposed triple-$\mathbf{q}$ ground state~\cite{Chen2021} features a vortex-like spin arrangement within hexagons, which are separated by alternating spins along the $c$ axis (Fig.~\ref{Fig_Intro}c). Despite being energetically degenerate with the zigzag configurations at certain high-symmetry points in the parameter space of $\mathcal{H}_\text{HK}$~\cite{Kruger2023}, we find that triple-$\mathbf{q}$ configurations generally have a higher classical energy than the zigzag ground states. In out-of-plane magnetic fields $\B\parallel c$, triple-$\mathbf{q}$ configurations are further energetically disfavored, as also shown in Ref.~\cite{Wang2023}. However, the inclusion of non-bilinear couplings -- such as the ring exchange interaction -- can stabilize a triple-$\mathbf{q}$ ground state.~\cite{Kruger2023,Wang2023,Francini2024PhaseDiag} Krüger~\textit{et al.} have shown that such a ring interaction can be included in linear spin wave theory by effective local fields and renormalization of bilinear couplings.~\cite{Kruger2023} Following this approach, we complement the extended Heisenberg-Kitaev Hamiltonian (Eq.~\ref{eq-HKHamiltonian}) by

\begin{equation}\label{eq_H-nbl}
    \mathcal{H}_\text{nbl} = - \sum_i h(\mathbf{n}_i\cdot\mathbf{S}_i)
\end{equation}

\noindent with the effective field strength $h$ to account for non-bilinear interactions. For the effective field directions $\mathbf{n}_i$, see the appendix, Sec.~\ref{Appendix_Sec_3q}. The effective local fields and the renormalization of the parameters in $\mathcal{H}_\text{HK}$ depend on the spin configuration and, therefore, may change with applied magnetic fields. Neglecting this potential field-dependence, we assume fixed parameters in $\mathcal{H}_\text{HK}$ and $\mathcal{H}_\text{nbl}$ in the following analysis. This assumption is an approximation for $B\neq0\,$T. For our spin wave calculation, the only reported \tripleq model for \NCTO \cite{Kruger2023} with $(J_1,K,\Gamma,\Gamma') = (1.23, -8.29, 1.86, -2.27)\,$meV, $(J_\text{2A},J_\text{2B},J_3) = (0.32, -0.24, 0.47)\,$meV and $h = 0.88\,$meV is used.

\begin{figure}[]
\includegraphics[width=\columnwidth,clip]{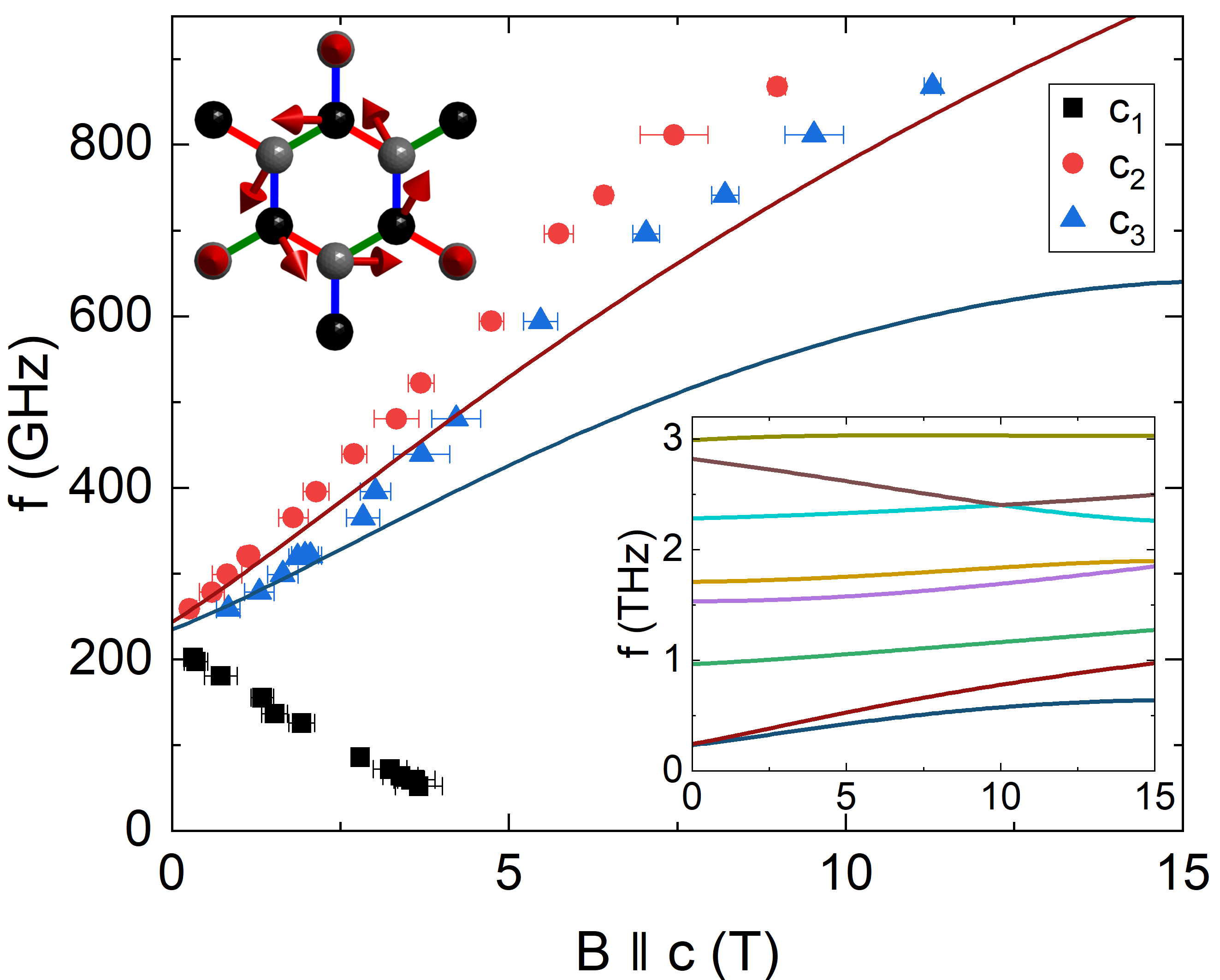}
\caption{Spin wave modes of the triple-$\mathbf{q}$ ground state in out-of-plane magnetic fields $\B\parallel c$. The spin wave frequencies (lines) calculated for the extended Kitaev-Heisenberg model with ring-exchange interaction implemented as effective local fields are displayed with the full spin wave spectrum of 8 modes in the lower-right inset. The experimental HF-ESR results at $T=2\,$K obtained in this work are shown as symbols. In the upper-left, the triple-$\mathbf{q}$ spin configuration at zero applied magnetic field is depicted. 
}
\label{Fig_SpinW3q}
\end{figure}

The so-calculated spin wave spectrum for the \tripleq ground state in out-of-plane magnetic fields $\B\parallel c$ is shown in Fig.~\ref{Fig_SpinW3q}. Among the eight calculated modes, two fall within the frequency range accessible by our HF-ESR experiment, while the remaining six modes lie at higher frequencies ($>950\,$GHz), as shown in the inset. Comparison with the experimentally observed three spin wave modes yields two main findings: The rising modes $c_2$ and $c_3$ are excellently predicted at low magnetic fields. In particular, their split zero-field gap structure (shown in detail in the inset of Fig.~\ref{Appendix_Sec_3q}) and the positive effective \textit{g}-factors are reproduced. We find that both characteristics -- split gap and the non-zero effective \textit{g}-factors -- are associated with different next-neighbor interactions $J_\text{2A}\neq J_\text{2B}$ on the two sublattices (see Fig.~\ref{Appendix_SpinWaveCalc_tripleQ}), which also rationalize the weak ferrimagnetism of \NCTO in the \tripleq scenario~\cite{Yao2020,Francini2024fM}. At intermediate fields, the measured and calculated spin wave frequencies begin to deviate. These deviations could be resolved by a minor adjustment of the parameters in $\mathcal{H}_\text{HK}$ and $\mathcal{H}_\text{nbl}$. Alternatively, they may be explained by a small magnetic field-dependence of the interactions: either physical as a consequence of spin-lattice coupling or due to parameter renormalization (see below Eq.~\eqref{eq_H-nbl}). 

The softening spin wave mode $c_1$, on the other side, is not predicted by the \tripleq model. It could be speculated that one of the high-energetic spin excitations is completely misplaced and instead should describe $c_1$ at low energies. However, this scenario is highly unlikely because it requires a reordering of the spin wave modes while still fitting to the modes $c_2,c_3$ and the INS data~\cite{Yao2022,Kruger2023}. Excluding such a mode reordering renders $c_1$ an \emph{additional spin wave mode} observed beyond the \tripleq model's excitations. The number of spin wave modes of a magnetically ordered system is determined by the number of moments in the magnetic supercell, i.e., the number of magnetic sublattices. An additional mode, therefore, implies a magnetic structure with a larger supercell. Hence, we conclude that the \tripleq ground state model for \NCTO is not yet complete.

\subsection{The role of interlayer coupling}
\label{Sec_InterlayerCoupling}

\noindent
Our comparison of the experimental data with the predicted spin wave modes yields the following implications: (1) The observation of an additional spin wave mode $c_1$ requires a larger magnetic structure than assumed in the current \tripleq model. (2) However, the field-dependent behavior of the modes $c_2,c_3$ is quantitatively correctly captured by the \tripleq model obtained from zero-field INS. Combining (1) and (2), this suggests that the \tripleq framework remains largely valid but requires expansion. A potential in-plane superstructure containing the $(2\times2)$ \tripleq magnetic unit cell (see Fig.~\ref{Fig_Intro}c) spreads over at least 16 magnetic sides for a $(2\times4)$ cell, while a more symmetric $(4\times4)$ supercell is even larger. It is unclear which interactions might stabilize such a large spin superstructure. Instead, an expansion from the up-to-now two-dimensional \tripleq model \cite{Kruger2023} to the full three-dimensional structure of \NCTO seems more reasonable. Including both hexagonal Co-layers in alternating stacking (see Fig.~\ref{Fig_Intro}a) provides 16 spin wave modes to describe HF-ESR and INS data. Introducing interlayer interactions, which need to couple Co-spins across the separating disordered sodium layers, will shift the frequencies of spin wave modes. If such a shift allows the spin wave spectrum to be refined so that the additional mode $c_1$ is reproduced, this will imply that the field-induced phase transition revealed by the softening of $c_1$ is associated with the interplane nature of magnetic order.

Revisiting the zigzag ground state models, we note that adding interlayer coupling doubles the number of spin wave modes from four to eight. However, such an interaction between the hexagonal planes is required to be small due to the vanishing spin wave dispersion for out-of-plane momenta~\cite{Chen2021}. Therefore, it only slightly splits the previously degenerate four modes and leads to four closely spaced pairs in the zero-field spin wave spectrum. Two of these pairs are required to describe the three modes $c_1,c_2$, and $c_3$ at low energy observed in this work. This leaves only two remaining close-lying pairs of spin wave modes that are insufficient to describe the five distinct excitations at the $\Gamma$-point and energies $>\Delta,\Delta_2$ in the INS data.~\cite{Yao2022,Chen2021} Thus, under the constraint of small interlayer coupling, zigzag ground states are excluded as viable models for Na\ti{2}Co\ti{2}TeO\ti{6}.

\section{Conclusions}

\noindent
In summary, we probed the low-energy spin excitations in the Kitaev material \NCTO by high-frequency/high-field electron spin resonance. For magnetic fields applied along the armchair direction $a^*$, several excitation modes were found in the three phases appearing in the studied field range together with a frequency-independent feature at the phase boundary associated with the putative quantum critical endpoint. Our central experimental result is the observation of three spin wave modes $c_1,c_2$, and $c_3$ observed for out-of-plane magnetic fields $\B\parallel c$. They exhibit a split excitation gap structure at zero magnetic field with two distinct gaps of $\Delta=211\,$GHz and $\Delta_2=237\,$GHz. The softening of the spin wave mode $c_1$ reveals a field-induced phase transition at $\Bcone = 4.7\,$T coinciding with a kink in isothermal magnetization.

Based on our spin wave calculations for $\B\parallel c$, the experimental data of this work exclude a zigzag ground state for Na\ti{2}Co\ti{2}TeO\ti{6}. The only \tripleq ground state model reported so far~\cite{Kruger2023} correctly describes the spin wave modes $c_2$ and $c_3$ but fails to reproduce the mode $c_1$. This implies a magnetic structure with larger magnetic unit cell and shows that the present \tripleq ground state picture is not complete. The search for a full spin model for the low-field antiferromagnetic phase of \NCTO is, thus, still open -- as it is for the phases at intermediate fields. To find a comprehensive model, we propose a combined analysis of high-quality single-crystal inelastic neutron scattering data \cite{Yao2020} at zero field together with the in-field behavior of spin waves reported in this work (e.g., as in Tab.~\ref{Tab_g-eff}). Such an analysis can be based on a non-approximate implementation of non-bilinear interactions and should include -- as suggested by our analysis --  different next-neighbor couplings $J_\text{2A}$ and $J_\text{2B}$ on the two crystallographic sublattices as well as interlayer interactions.

\begin{acknowledgments}
We thank Alexey Alfonsov for valuable discussions.
Support by Deutsche  For\-schungs\-ge\-mein\-schaft (DFG) under Germany’s Excellence Strategy EXC2181/1-390900948 (The Heidelberg STRUCTURES Excellence Cluster) is gratefully acknowledged. L.B. acknowledges funding by the German Research Foundation (DFG) via Research Training Group GRK 2948/1. J.A. acknowledges support by the IMPRS-QD Heidelberg. KYC was supported by the National Research Foundation (NRF) of Korea (Grant Nos.~2020R1A5A1016518 and RS-2023-00209121). R.S.~acknowledges the financial support provided by the Ministry of Science and Technology in Taiwan under project numbers NSTC 111-2124-M-001-007, Financial support from the Center of Atomic Initiative for New
Materials (AI-Mat), (Project No.~108L9008) and Academia Sinica for the budget of AS-iMATE-113-12.

\end{acknowledgments}

\bibliography{NCTO-ESR}

\hfill

\appendix
\section*{Appendix}

\renewcommand{\thefigure}{A\arabic{figure}}
\setcounter{figure}{0}  

\section{Spin wave calculations for Na\ti{2}Co\ti{2}TeO\ti{6}}
\label{Sec_Appendix}

\begin{figure}[]
\includegraphics[width=\columnwidth,clip]{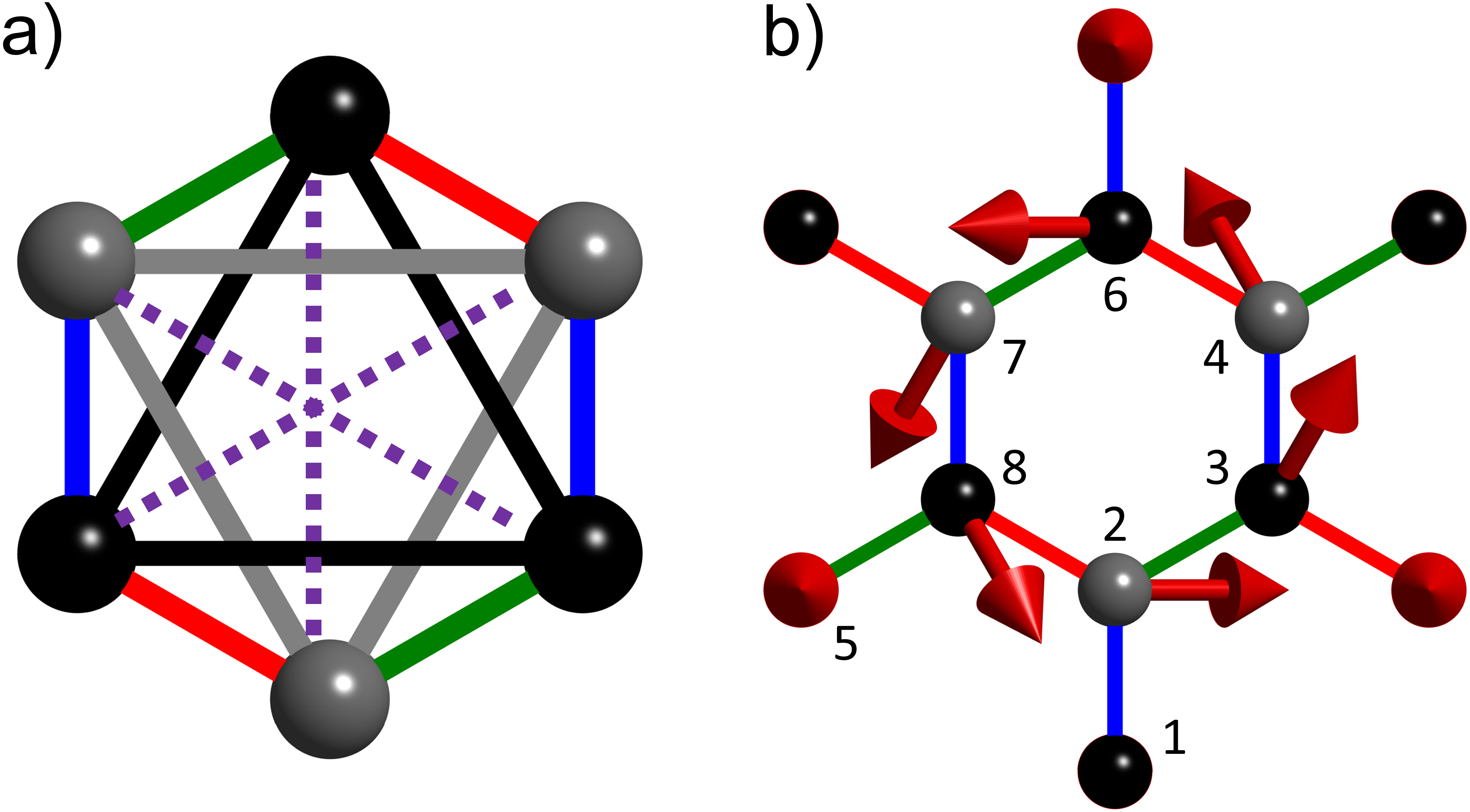}
\caption{(a) Scheme of the Heisenberg-Kitaev Hamiltonian $\mathcal{H}_\text{HK}$ showing the nearest-neighbor $x$-bonds (green), $y$-bonds (red) and $z$-bonds (blue), the next-neighbor Heisenberg couplings $J_\text{2A}$ and $J_\text{2B}$ on the Co-sublattices A (black) and B (gray), respectively, and the third-neighbor coupling $J_3$ (purple, dashed). (b) The hexagonal magnetic unit cell of the \tripleq structure with sites enumerated as in Eq.~\eqref{Appendix_eq_Hnbl-directions}. On site~1, the spin points antiparallel to the $c$ axis and is obscured by the ion.}
\label{Appendix_SpinWaveCalc}
\end{figure}

\noindent Spin wave frequencies were calculated using linear spin wave theory for the Heisenberg-Kitaev Hamiltonian $\mathcal{H}_\text{HK}$ defined in Eq.~\eqref{eq-HKHamiltonian}. The interactions of $\mathcal{H}_\text{HK}$ introduced in the main text are depicted in Fig.~\ref{Appendix_SpinWaveCalc}a.
Their interaction strengths in all spin models proposed for Na\ti{2}Co\ti{2}TeO\ti{6} are summarized in Table~\ref{Appendix_Tab_H-HK-parameter}.
An applied magnetic field $\mathbf{B}$ is included by the Zeeman term 
\begin{equation}
    \mathcal{H}_\text{Z}= \muB\sum_i\mathbf{S}_i^\top g \mathbf{B}
\end{equation}
with an anisotropic \textit{g}-tensor with $g_c = 2.3$ and $g_{ab} = 4.13$ found by ESR measurements at $T=50\, $K~\cite{Lin2021}. Since the calculations are performed for out-of-plane magnetic fields $\mathbf{B} = B \times \mathbf{e}_c$, only the $c$-component of the $g$-tensor enters the calculation due to $\muB\mathbf{S}_i^\top g \mathbf{B} = \muB g_c B\times S_i^c$.

\begin{figure}[]
\includegraphics[width=\columnwidth,clip]{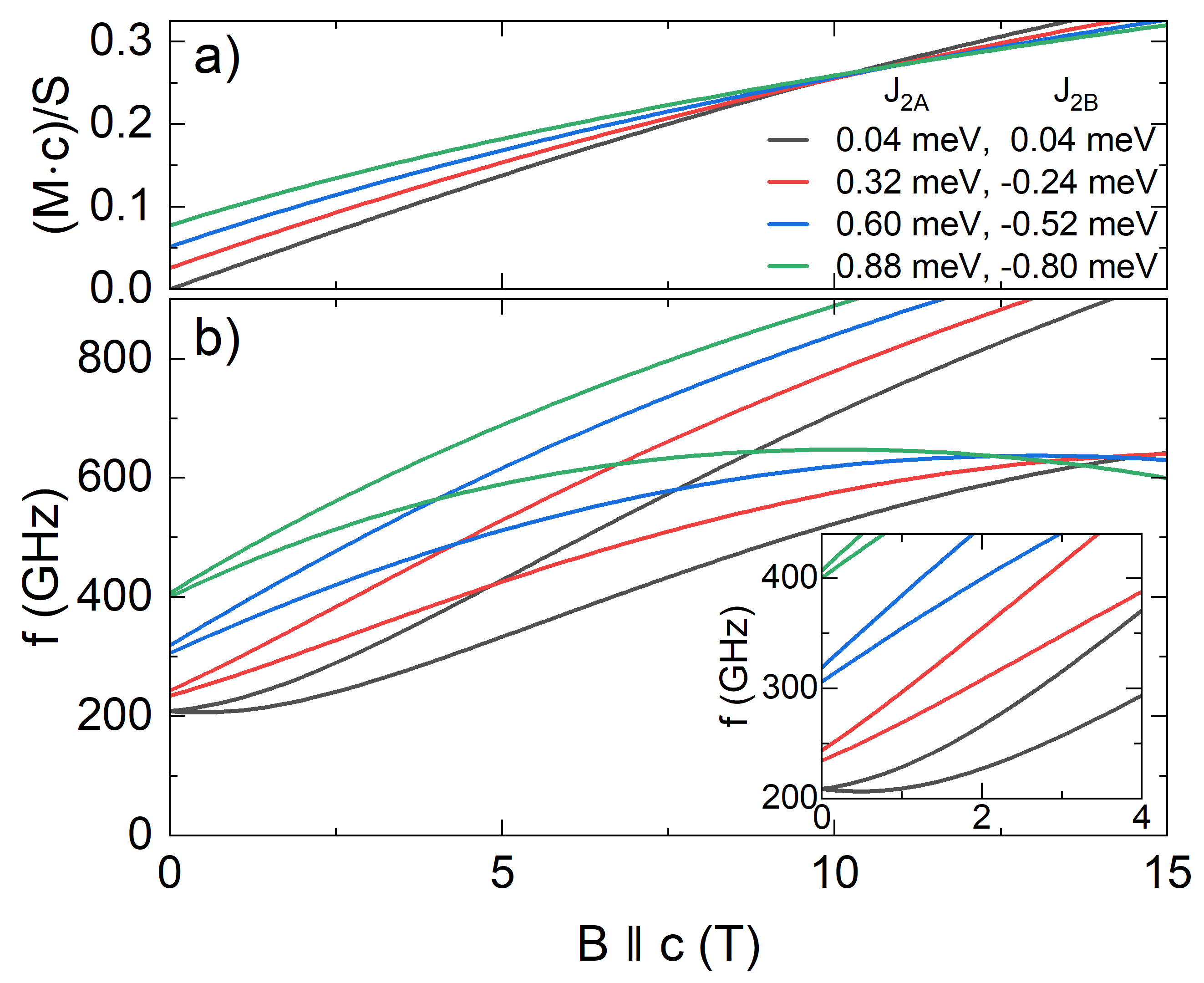}
\caption{Characteristic quantities of the triple-$\mathbf{q}$ ground state in an applied magnetic field $\B\parallel c$ for several differences of the next-neighbor interactions $J_\text{2A}-J_\text{2B}$ on the two sublattices. The difference $J_\text{2A}-J_\text{2B} = 0.56\,\text{meV}$ (red) of the \tripleq model is compared to vanishing (gray) and larger differences of $1.12\,\text{meV}$ (blue) and $1.68\,\text{meV}$ (green). a) Normalized magnetization along the $c$ axis shows the ferrimagnetic moment induced by $J_\text{2A}-J_\text{2B}\neq 0$. (b) Spin wave frequencies at low energies with the inset highlighting the behavior at low fields.}
\label{Appendix_SpinWaveCalc_tripleQ}
\end{figure}

\subsection{Zigzag ground state models}

\label{Appendix_SecSW_Zigzag}
\begin{figure*}[p]
\includegraphics[width=\textwidth,clip]{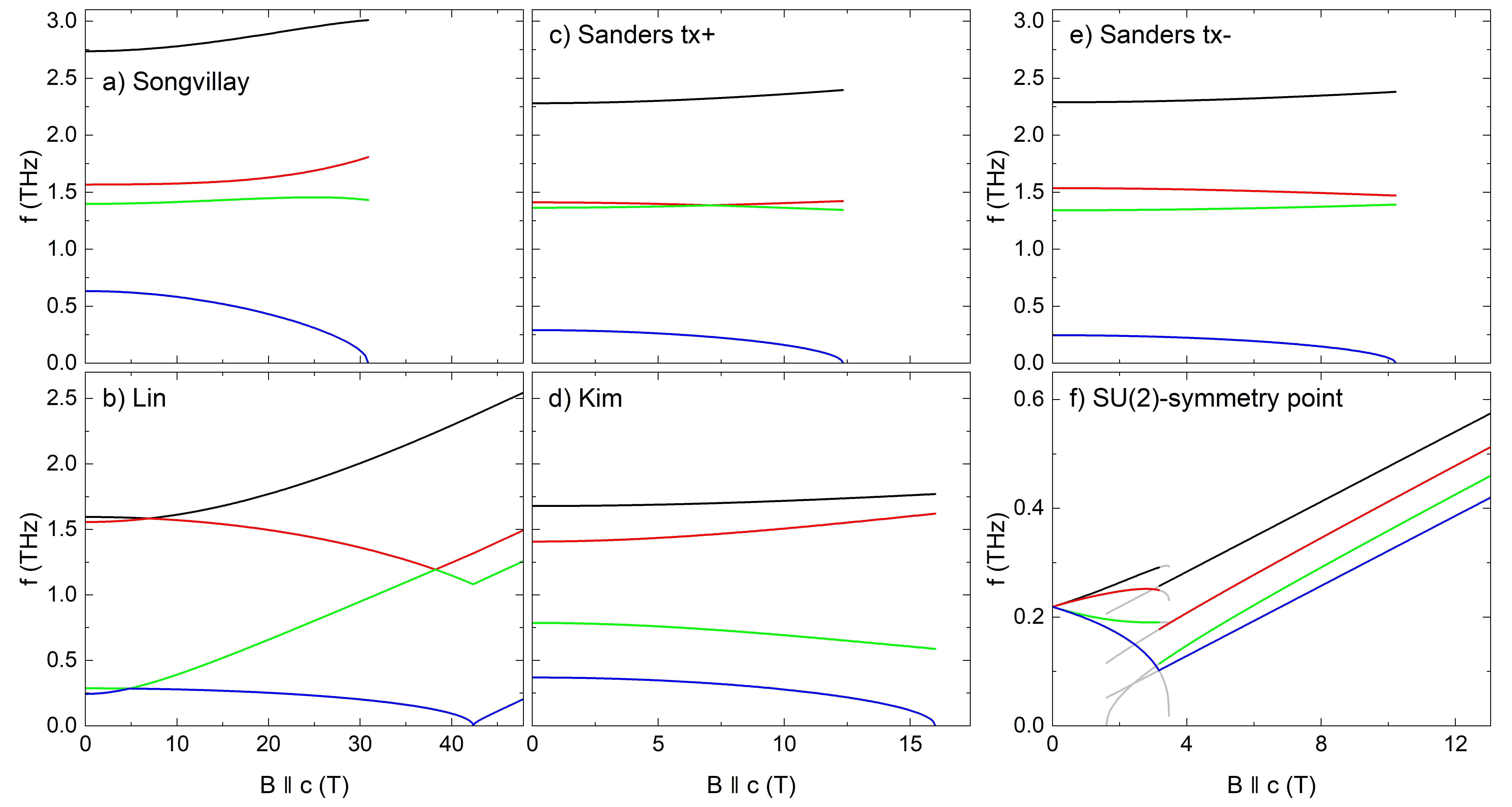}
\caption{Spin wave frequencies for models which exhibit the zigzag ground state at zero field. The parameters are proposed by (a) Songvillay~\textit{et al.}~\cite{Songvilay2020}, (b) Lin~\textit{et al.}~\cite{Lin2021}, (c,e) Sanders~\textit{et al.}~\cite{Sanders2022} and (d) Kim~\textit{et al.}~\cite{Kim2021}. The zigzag ground states are stable at low fields up to the softening of the lowest mode (blue). In (b), the spin wave frequencies are also shown for the directly adjacent polarized state above $42\,$T. (f) The spin wave frequencies for the $\mathcal{H}_\text{HK}$ parameters at the SU(2)-symmetric point~\cite{Kruger2023} are shown for the zigzag ground state at low fields and the polarized state at high fields. The parameters are scaled to fit the experimentally observed excitation gap. Gray lines depict the spin wave frequencies for fields where the respective state is metastable but not the ground state.}
\label{Appendix_SpinWaveCalc_zigzag}
\end{figure*}

\noindent The spin wave frequencies of zigzag ground state models for \NCTO reported in Refs.~\cite{Songvilay2020,Lin2021,Sanders2022,Kim2021} are shown in Fig.~\ref{Appendix_SpinWaveCalc_zigzag}a-e. In the zigzag phase at low fields, all reported models exhibit the three characteristic properties discussed in the main text: (1) Only one or two spin wave modes at low frequencies close to the excitation gap $\Delta = 211\,$GHz; (2) flat frequency-field dependence of all modes at low magnetic fields; (3) softening of the energetically lowest spin wave mode and phase transition to the adjacent phases only at fields larger than $10\,$T. With this, all these models are incompatible with the spin wave modes observed in this work and cannot be realized in Na\ti{2}Co\ti{2}TeO\ti{6}. 

The zigzag models by Samarakoon \textit{et al.}~\cite{Samarakoon2021} include a small interplane coupling (see Tab.~\ref{Appendix_Tab_H-HK-parameter}). Neglecting this, the spin wave spectra of models A  and B are quantitatively similar to the models tx+ \cite{Sanders2022} shown in Fig.~\ref{Appendix_SpinWaveCalc_zigzag}c and thus incompatible with our data as well. Including the proposed interlayer coupling cannot resolve this incompatibility, as discussed in Sec.~\ref{Sec_InterlayerCoupling} of the main text.

In Fig.~\ref{Appendix_SpinWaveCalc_zigzag}f, the spin wave frequencies for the zigzag ground state at the SU(2)-symmetric point are shown. The interactions in $\mathcal{H}_\text{HK}$ are globally scaled to fit the experimentally observed excitation gap at zero magnetic field. This parameter set was not proposed as a model for \NCTO but as a high-symmetry point in the parameter space proximate to the real interaction parameters of the compound.~\cite{Kruger2023} Since all spin wave modes are degenerate at zero magnetic field and thus incompatible with experimental results, the SU(2)-symmetric point parameter set is not considered in our analysis of models for Na\ti{2}Co\ti{2}TeO\ti{6}. However, the SU(2)-symmetric point is of theoretical interest because it exhibits a degenerated ground state manifold that contains the zigzag configurations and the \tripleq state.

\begin{table*}[p]
    \centering
    \def\arraystretch{1.5}
    \caption{Parameters of the extended Heisenberg-Kitaev models reported in literature for Na\ti{2}Co\ti{2}TeO\ti{6}. All interactions are given in meV. The models in the first block exhibit a zigzag ground state at zero magnetic field. At the SU(2)-symmetric point, the zigzag and \tripleq configurations are degenerate in energy. The model by Krüger~\textit{et al.} stabilizes a \tripleq ground state.  Some models include additional interactions (interplane interaction $J_c$, local effective field $h$) as noted in the last column.\\}
    \begin{tabular}{L{0.18\textwidth}|C{0.08\textwidth}C{0.08\textwidth}C{0.08\textwidth}C{0.08\textwidth}|C{0.08\textwidth}C{0.08\textwidth}C{0.08\textwidth}|C{0.15\textwidth}}
                                     & $J_1$  & $K$& $\Gamma$&$\Gamma'$& $J_\text{2A}$ & $J_\text{2B}$ & $J_3$  &   \\
     \hline
    Songvillay \cite{Songvilay2020}  & -0.1   & -9.0   & 1.8   & 0.3   & \multicolumn{2}{c}{0.3}    &    0.9 &     \\ 
    Kim \cite{Kim2021}               & -1.5   & 3.3    & -2.8  & 2.1   & \multicolumn{2}{c}{--}     &    1.5 &     \\ 
    Lin \cite{Lin2021}               & -2.175 & 0.125  & 0.125 &  --     & \multicolumn{2}{c}{--}     &    2.5 &     \\ 
    Samarakoon A \cite{Samarakoon2021}& -0.2  & -7.0   & 0.02 & -0.23 & \multicolumn{2}{c}{0.05}   &    1.2 &  $J_c = -0.15\,$meV  \\ 
    Samarakoon B \cite{Samarakoon2021}& -3.2  & 2.7    & -2.9  & 1.6   & \multicolumn{2}{c}{0.1}    &    1.2 &  $J_c = -0.4\,$meV   \\
    Sanders tx+ \cite{Sanders2022}    & -3.5  & 3.2    & -3.0  & 2.0   & \multicolumn{2}{c}{0}      &    1.4 &     \\ 
    Sanders tx- \cite{Sanders2022}    & -0.2  & -7.0   & 0.5   & 0.15  & \multicolumn{2}{c}{0}      &    1.6 &  \\ \hline 
    SU(2)-sym. point \cite{Kruger2023}&$-1/9$&$-2/3$&$8/9$& $-4/9$& \multicolumn{2}{c}{0}      &    0   &  \\ \hline 
    Krüger \cite{Kruger2023}             & 1.23  & -8.29  & 1.86  & -2.27 & 0.32       &  -0.24         &     0.47&  $h = 0.88\,$meV     \\ \hline 
    \end{tabular}
    \label{Appendix_Tab_H-HK-parameter}
\end{table*}

\subsection{Triple-$\mathbf{q}$ ground state model}
\label{Appendix_Sec_3q}

The studied \tripleq configurations (see Sec.~\ref{Sec_Methods}) are generally not energetically minimal states of the Kitaev-Heisenberg-model and exhibit higher classical energy than the zigzag ground states. For example, assuming the interaction parameters of the tx+ (tx-) model, the classical energy of the optimal \tripleq configuration is $-1.483\,$meV/spin ($-1.505\,$meV/spin), while the zigzag ground state exhibits $-1.516\,$meV/spin ($-1.529\,$meV/spin). However, non-bilinear interactions can stabilize the \tripleq structure as the ground state of the extended Heisenberg-Kitaev Hamiltonian $\mathcal{H}_\text{HK}$. For our calculation, we employ the model by Krüger~\textit{et al.}~\cite{Kruger2023}, which implements these higher-order interactions in linear spin wave theory by effective local fields and a renormalization of the interaction parameters. The effective local field term $\mathcal{H}_\text{nbl}$ for the spins of the \tripleq structure is defined in Eq.~\eqref{eq_H-nbl}.
The directions $\mathbf{n}_i$ of the effective local fields point along the spin directions of the \tripleq ground state of $\mathcal{H}_\text{HK}$ at the SU(2)-symmetric point (see Tab.~\ref{Appendix_Tab_H-HK-parameter} and Sec.~\ref{Appendix_SecSW_Zigzag}). They are given by

\begin{equation}\label{Appendix_eq_Hnbl-directions}
	\Big(\,\mathbf{n}_i\,\Big)_{i=1\dots4} = 
	\begin{pmatrix}
		0  & \frac{2\sqrt{2}}{3} & \frac{\sqrt{2}}{3} & -\frac{\sqrt{2}}{3} \\[6pt]
		0  & 0                   & \sqrt{\frac{2}{3}} & \sqrt{\frac{2}{3}}  \\[6pt]
		-1 & -\frac{1}{3}        & \frac{1}{3}        & -\frac{1}{3}
	\end{pmatrix}
	=
	-\Big(\,\mathbf{n}_i\,\Big)_{i=5\dots8}
\end{equation}\\

\noindent with the site labeling as denoted in Fig.~\ref{Appendix_SpinWaveCalc}b. The complete Hamiltonian used in the calculations is then given by $\mathcal{H}_\text{HK}+\mathcal{H}_\text{Z}+\mathcal{H}_\text{nbl}$.  The spin wave spectrum for the \tripleq model is shown in Fig.~\ref{Fig_SpinW3q} in the main text. We found the symmetry-breaking of the next-neighbor Heisenberg coupling $J_\text{2A}\neq J_\text{2B}$ on the two Co-sublattices to have a relevant effect on the spectra: Calculations for different splittings $J_\text{2A}-J_\text{2B}$ but fixed average $(J_\text{2A}+J_\text{2B})/2 = 0.04\,$meV of the two couplings reveal that this splitting leads to the split excitation gap structure as well as the non-zero effective $g$-factors (slope) of the spin wave modes at zero field as shown in Fig.~\ref{Appendix_SpinWaveCalc_tripleQ}. Furthermore, $J_\text{2A}\neq J_\text{2B}$ generates a ferrimagnetic moment of the \tripleq structure by unequal out-of-plane canting of the vertex hexagon spins (2-4,6-8 in Fig.~\ref{Appendix_SpinWaveCalc}b) on the two different Co-sublattices.

\end{document}